\documentclass{iaus}
\usepackage{graphicx}
\usepackage{psfig}
\newcommand\la{\raisebox{-0.5ex}{$\,\stackrel{<}{\scriptstyle\sim}\,$}}
\newcommand\ga{\raisebox{-0.5ex}{$\,\stackrel{>}{\scriptstyle\sim}\,$}}
\newcommand{\lya}{\ifmmode {\rm Ly}\alpha \else Ly$\alpha$\fi}
\newcommand{\hii}{H~{\sc ii}}
\def\msun{\ifmmode M_{\odot} \else M$_{\odot}$\fi}
\def\msunyr{\ifmmode M_{\odot} {\rm yr}^{-1} \else M$_{\odot}$ yr$^{-1}$\fi}
\def\zsun{\ifmmode Z_{\odot} \else Z$_{\odot}$\fi}
\def\lsun{\ifmmode L_{\odot} \else L$_{\odot}$\fi}
\def\micron{$\mu$m}

\def\mup{\ifmmode M_{\rm up} \else M$_{\rm up}$\fi}
\def\mlow{\ifmmode M_{\rm low} \else M$_{\rm low}$\fi}
\def\aap{A\&A}

\def\aj{AJ}
\def\apj{ApJ}
\def\apjl{ApJ}

\def\mnras{MNRAS}

\def\vwfpc{V$_{\rm 606W}$}
\def\iwfpc{I$_{\rm 814W}$}
\def\zacs{z$_{\rm 850LP}$}
\def\jnic{H$_{\rm 110W}$}
\def\hnic{H$_{\rm 160W}$}

\def\flyf{\ifmmode f_{\rm Lyf} \else $f_{\rm Lyf}$\fi}
\def\pz{\ifmmode P(z) \else $P(z)$\fi}
\def\ki2{\ifmmode \chi^2 \else $\chi^2$\fi}
\def\zphot{\ifmmode z_{\rm phot} \else $z_{\rm phot}$\fi}
\title[$z \ga 6$ galaxies]{Stellar populations and \lya\ emission from lensed
$z \ga 6$ galaxies}

\author[D.\ Schaerer \& R.\ Pell\'o]%
{Daniel Schaerer$^{1,2}$,
\and Roser Pell\'o$^2$}

\affiliation{$^1$Observatoire de Gen\`eve,
51, Ch. des Maillettes, CH-1290 Sauverny, Switzerland
 email: daniel.schaerer@obs.unige.ch\\[\affilskip]
$^2$Observatoire Midi-Pyr\'en\'ees,
14 Avenue E. Belin, F-31400 Toulouse, France email: roser@ast.obs-mip.fr}

\pubyear{2004}
\volume{225}
\pagerange{1--8}
\date{?? and in revised form ??}
\jname{Impact of Gravitational Lensing on Cosmology}
\editors{Mellier, Y. \& Meylan,G. eds.}
\begin{document}

\maketitle

%%%%%%%%%%%%%%%%%%%%%%%%%%%%%%%%%%%%%%%%%%%%%%%%%%%%%%%%%%%%%%%%%%%%%%%%
\begin{abstract}
We present results from an SED analysis of two lensed high-$z$ objects,
the $z=6.56$ galaxy HCM6A behind the cluster Abell 370 discovered by 
Hu \etal\ (2002) and the triple arc at $z \sim 7$ behind Abell 2218
found by Kneib \etal\ (2004).
For HCM 6A we find indications for the presence of dust in
this galaxy, and we estimate the properties of its stellar populations
(SFR, age, etc.), and the intrinsic \lya\ emission. 
From the ``best fit'' reddening ($E(B-V) \sim 0.25$) its estimated luminosity
is $L \sim (1-4) \times 10^{11} \lsun$, in the range of luminous infrared galaxies.
For the arc behind Abell 2218 we find a most likely redshift of $z \sim$ 6.0--7.2
taking into account both our photometric determination and lensing considerations.
SED fits indicate generally a low extinction 
but do not strongly constrain the SF history. 
Best fits have typical ages of $\sim$ 3 to 400 Myr.
The apparent 4000 \AA\ break observed recently by Egami \etal (2004) from 
combination of IRAC/Spitzer and HST observations can also well be 
reproduced with templates of young populations ($\sim$ 15 Myr or even younger)
and does not necessarily imply old ages.
Finally, we briefly examine the detectability of dusty lensed high-z galaxies with
Herschel and ALMA. 
\end{abstract}

%%%%%%%%%%%%%%%%%%%%%%%%%%%%%%%%%%%%%%%%%%%%%%%%%%%%%%%%%%%%%%%%%%%%%%%%
\section{Introduction}

Little is known about the stellar properties, extinction, and
the expected intrinsic \lya\ emission of distant, high redshift 
galaxies. Indeed, although it has in the recent past become possible 
through various techniques to detect already sizeable numbers of 
galaxies at $z \ga 5$ 
(see e.g.\ the reviews of Taniguchi et al.\ 2003 and Spinrad 2003)
the information available on these objects remains generally scant.
For example, in many cases the galaxies are just detected in
two photometric bands and \lya\ line emission, when present, 
serves to determine the spectroscopic redshift (e.g.\ Bremer \etal\ 2004,
% Bouwens \etal\ 2004, --> has more than 2 bands !
Dickinson \etal\ 2004, Bunker \etal\ 2004).
Then the photometry is basically used to estimate the star formation rate 
(SFR) assuming standard conversion factors between the UV restframe
light and the SFR, and nothing is known about the extinction,
and the properties of the stellar population (such as age, detailed 
star formations history etc.) 

At higher redshift ($z \ga 6$) even less information is generally available.
Many objects are found by \lya\ emission, but remain weak or sometimes even 
undetected in the continuum (e.g.\ Rhoads \& Malhotra 2001, Kodaira \etal\ 2003,
Cuby \etal\ 2003, Ajiki \etal\ 2003, Taniguchi et al.\ 2004).
In these cases the \lya\ luminosity can be determined 
and used to estimate a SFR using again standard conversion factors. 
Also the \lya\ equivalent width is estimated,
providing some possible clue on the nature of these source.
However, this has lead to puzzling results e.g.\ for the 
sources from the LALA survey (Malhotra \& Rhoads 2001, Rhoads \etal\ 2003) 
leaving largely open the question of the nature of these objects, their
stellar populations, extinction etc.

Strong gravitation lensing is extremely ``helpful'' for a large number
problems discussed at this conference, including also the present one.
In particular strong lensing has allowed to detect several of the
highest redshift galaxies known today (e.g.\ Ellis \etal\ 2001,
Hu \etal\ 2002, Kneib \etal\ 2004, Pell\'o \etal\ 2004a,
and the review of Pell\'o \etal\ 2003).
Also, thanks to the lensing magnification, it has been possible
to obtain photometric observations of reasonable quality in several bands
for some of these objects. For example it has even very recently been possible
to image a $z \sim 7$ galaxy with the Spitzer observatory at 3.6 and 4.5 \micron\
(Egami \etal\ 2004) !
As we'll show below (Sects.\ 2 and 3) this allow us to perform a quantitative SED
analysis to constrain properties  of the stellar populations, such as age and star 
formation (hereafter SF) history (burst or constant SF?), their extinction,
intrinsic \lya\ emission etc.
A detailed account of this work will be published elsewhere 
(Schaerer \& Pell\'o 2004).

As such, gravitational lensing provides a unique opportunity to
learn more about some selected high-$z$ galaxies. 
If generalised and applied to larger samples in the near
future, systematic studies of the properties of lensed high-$z$ galaxies 
could provide unique insights and complementary information to 
other deep/ultra-deep surveys targetting blank fields.
Also, extensions to wavelengths beyond the optical and near-IR
with existing facilities (e.g.\ in the radio, mm, and possibly sub-mm)
and future observatories should be of great interest, as briefly outlined 
for Herschel and ALMA in Sect.\ 4. 

%%%%%%%%%%%%%%%%%%%%%%%%%%%%%%%%%%%%%%%%%%%%%%%%%%%%%%%%%%%%%%%%%%%%%
\section{Stellar populations and dust in a lensed $z=6.56$ starburst galaxy}
The lensed $z=6.56$ galaxy HCM6A was found by Hu et al.\ (2002)
from a narrow-band survey in the field of the lensing cluster Abell 370.
Its redshift is established from the broad-band SED including a strong
spectral break, and from the observed asymmetry of the detected emission
line identified as \lya.

We have recently analysed the SED of this object by means of quantitative
SED fitting techniques using a modified version of the {\em Hyperz} code of 
Bolzonella et al.\ (2000)
\footnote{To convert the observed/adjusted quantities to absolute values
we adopt the following cosmological parameters:
$\Omega_m=0.3$, $\Omega_\Lambda=0.7$, and 
$H_0 = 70$ km s$^{-1}$ Mpc$^{-1}$.}.
The observed $VRIZJHK^\prime$ data are taken from Hu et al.\ (2002).
The gravitational magnification of the source is $\mu=4.5$
according to Hu et al.
The main free parameters of the SED modeling are
the spectral template, extinction, and the reddening law.
Empirical and theoretical templates including in particular 
starbursts and QSOs (SB+QSO templates), and predictions from synthesis models of 
Bruzual \& Charlot (BC+CWW group) and from Schaerer (2003, hereafter S03) are used. 

\begin{figure*}[htb]
%\centerline{\psfig{figure=plot_sed_6a.eps,width=7cm}
%	    \psfig{figure=plot_sed_6a_csfr_sbs.eps,width=7cm}}
\centerline{\psfig{figure=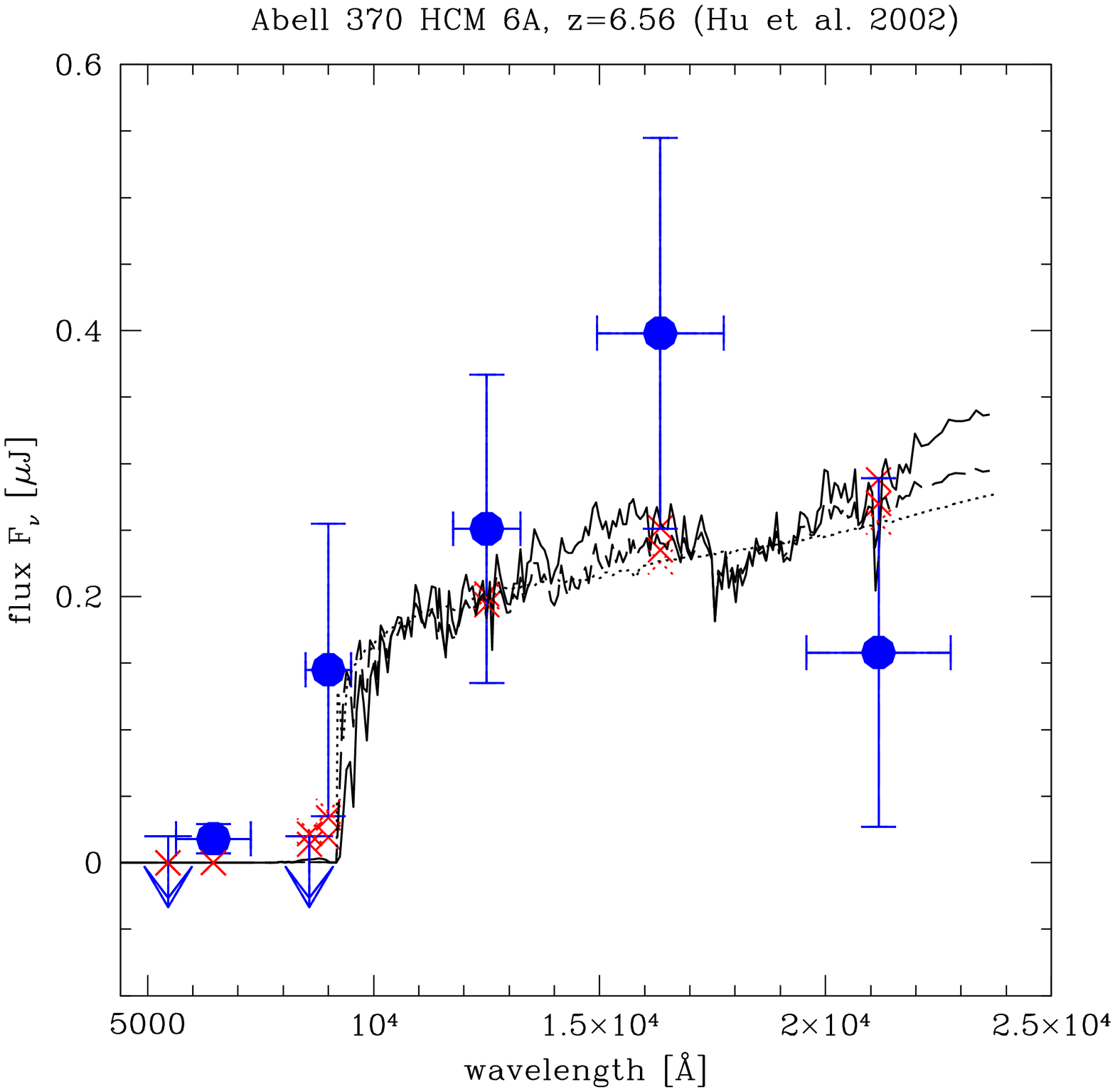,width=7cm}
	    \psfig{figure=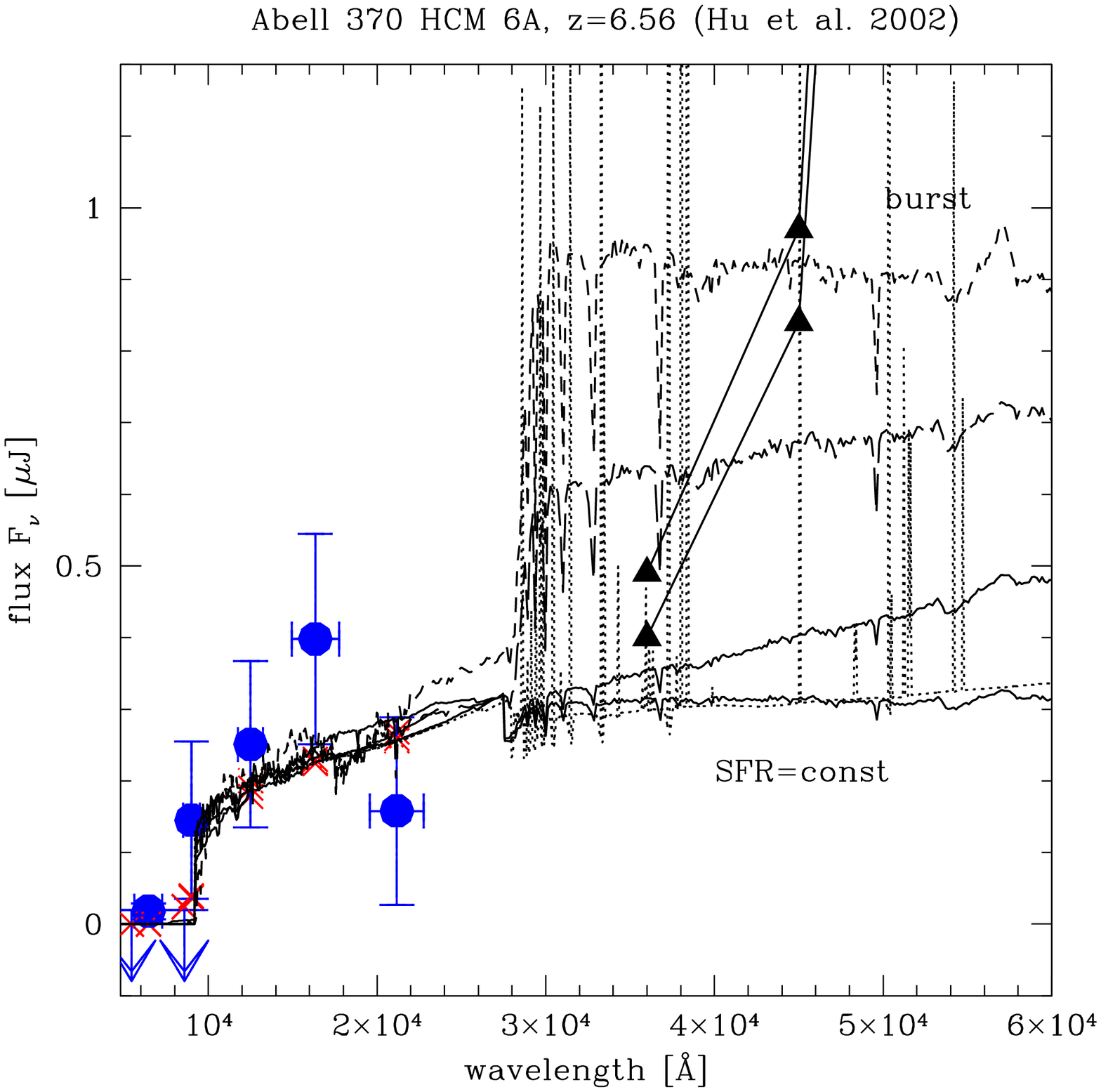,width=7cm}}
\caption{Best fits SEDs to the observations of Abell 370 HCM 6A.
The red crosses indicate the corresponding model broad band fluxes.
Solid lines show the best fit for a template from the BC+CWW group,
dotted from SB+QSO group, and dashed from the S03+ group 
%(see explanations in Sect.\ \ref{s_models}). 
{\bf Left:} Observed spectral range.
{\bf Right:} Predicted SED in Spitzer/IRAC domain for best fit models.
Dashed lines show the bursts from the BCCWW and S03+ template groups.
The dotted line 1is the spectrum of SBS 0335-052 from the SB+QSO group
with additional $A_V=1.$ The solid lines show best fits for constant
star formation using different extinction/attenuation laws (Calzetti 
starburst law versus SMC law). The solid triangles illustrate
the IRAC point-source sensitivity (1 $\sigma$) for low and medium
backgrounds excluding ``confusion noise''.}
\label{fig_sed_6a}
\end{figure*}

Overall the SED of HCM 6A (see Fig.\ 1)
is ``reddish'', showing an increase of the flux from Z to H
and even to K\footnote{The significance of a change of the SED slope
between JH and HK seems weak, and difficult to understand.}.
From this simple fact it is already clear 
qualitatively that one is driven towards stellar populations
with a) ``advanced'' age and little extinction or b) constant or 
young star formation  plus extinction.
However, for HCM6A a) can be excluded as no \lya\ emission would be expected 
in this case.

Quantitatively, the best solutions obtained for three ``spectral template groups''
are shown in the left panel of Fig.\ 1.
The solutions shown correspond to bursts of ages $\sim$ 50--130 Myr
%(BCCWW+, S03 templates) 
and little or no extinction.
However, as just mentioned, solutions lacking young ($\la$ 10 Myr) massive stars can
be excluded since \lya\ emission is observed.
The best fit empirical SB+QSO template shown corresponds to the spectrum of 
a metal-poor starburst galaxy with an extinction of $A_V \sim 1.$
On the basis of the present observations a narrow line (type II) AGN cannot be ruled out.
To reconcile the observed SED with \lya, a young population
or constant SF is required. 
In any of these cases fitting the ``reddish'' SED requires
a non negligible amount of reddening.

Although all best fit models require reddening, this result
is at present indicative and need to be firmed up.
Quantitatively (e.g.\ for constant star formation, 
solar metallicity models, Calzetti law) 
$A_V$ is typically $\sim$ 0.5--1.8 mag at the 68 \% confidence
level. Also somewhat smaller extinction can be obtained if 
the steeper SMC extinction law of Pr\'evot et al.\ (1984) 
is adopted.
Zero extinction cannot be ruled out at the $\sim 2 \sigma$ level. 
Better photometric accuracy, especially in the JHK bands, is needed
to reduce the present uncertainties and hence confirm the 
indication for dust.

From the best fit constant SF models
we deduce an extinction corrected star formation rate of
the order of SFR(UV) $\sim$ 11 -- 41 \msunyr\ for a Salpeter
IMF from 1 to 100 \msun\ or a factor 2.55 higher for the often adopted
lower mass cut-off of 0.1 \msun.
For continuous SF over timescales $t_{\rm SF}$ longer than $\sim$ 10 Myr, the total
(bolometric) luminosity output is typically $\sim 10^{10}$ \lsun\ per unit
SFR (in \msunyr) for a Salpeter IMF from 1-100 \msun, quite independently of metallicity. 
The total luminosity  associated with the observed SF is therefore 
$L \sim (1-4) \times 10^{11} \lsun$, in the range of luminous infrared galaxies
(LIRG).
For $t_{\rm SF} \sim$ 10 Myr the estimated stellar mass is 
$M_\star \approx t_{\rm SF} \times SFR \sim (1-4) \times 10^8$ \msun. 
Other properties such as the ``\lya\ transmission'' can also
be estimated from this approach.
A relatively high \lya\ transmission of $\sim$ 20--50 \% but possibly
up to $\sim$ 90 \% is estimated from our best fit models
(see Schaerer \& Pell\'o 2004).

It is interesting to examine the SEDs predicted by the various
models at longer wavelengths, including the rest-frame optical
domain, which is potentially observable with the sensitive IRAC camera
onboard the Spitzer Observatory and other future missions.
In the right panel of Fig.\ 1 we plot 
again the 3 best fits.
We see that these solutions have fluxes comparable to or above 
the detection limit of IRAC/Spitzer 
\footnote{See {\tt http://ssc.spitzer.caltech.edu/irac/sens.html}}.
%\footnote{The IRAC detection limits plotted here
%correspond to the values given by the Spitzer Science Center on 
%{\tt http://ssc.spitzer.caltech.edu/irac/sens.html} as 1 $\sigma$ point-source sensitivity 
%for low and medium backgrounds for frame times of 200s and described by Fazio \etal\ (2004).
%These values do not include ``confusion noise''.}.
%
On the other hand the strongly reddened constant SF or young burst solutions
do not exhibit a Balmer break and are hence expected to show fluxes
just below the IRAC sensitivity at 3.6 \micron\ and significantly
lower at longer wavelengths.
As \lya\ emission is expected only for the reddened SEDs the
latter solutions are predicted to apply to HCM 6A.
If possible despite the presence of other nearby sources,
IRAC/Spitzer observations of HCM 6A down to the 
detection limit or observations with other future satellites 
could allow to verify our prediction and therefore provide an independent
(though indirect) confirmation of the presence of dust in this high-z
galaxy.

%%%%%%%%%%%%%%%%%%%%%%%%%%%%%%%%%%%%%%%%%%%%%%%%%%%%%%%%%%%%%%%%%%%%%
\section{A lensed galaxy at $z \sim$ 6--7 behind Abell 2218}
This interesting triply imaged object, a possible $z \sim 7$ galaxy, 
has recently been discovered by Kneib et al.\ (2004, hereafter KESR) 
from deep $Z$ band observations with ACS/HST. 
In the meantime it has also been observed with Spitzer
(see Richard \etal, these proceedings; Egami \etal\ 2004).
The currently available observations include 
\vwfpc (undetected), \iwfpc, \zacs, $J$, \jnic, \hnic, and 3.6 and 4.5 
\micron\ with IRAC/Spitzer.
The photometry from these authors has been adopted here to
analyse the properties of this object in a similar way as for
HCM 6A. In practice, small differences are found in the published
photometry; we therefore adopt three different SEDs (SED1-3) to
describe this object (see Schaerer \& Pell\'o 2004 for details).
No emission line has so far been detected for Abell 2218 KESR.
Its spectroscopic redshift remains therefore presently unknown
but the well-constrained mass model for the cluster strongly suggests 
a redshift $z \sim$ 6.5--7 for this source.
The magnification factors of both images a and b is $\mu=25 \pm 3$,
according to KESR.

\begin{figure*}[bt]
%\centerline{\psfig{figure=plot_pz_7_rev1.eps,width=7cm}
%	    \psfig{figure=plot_sed_7rev_oldyoung.eps,width=7cm}}
\centerline{\psfig{figure=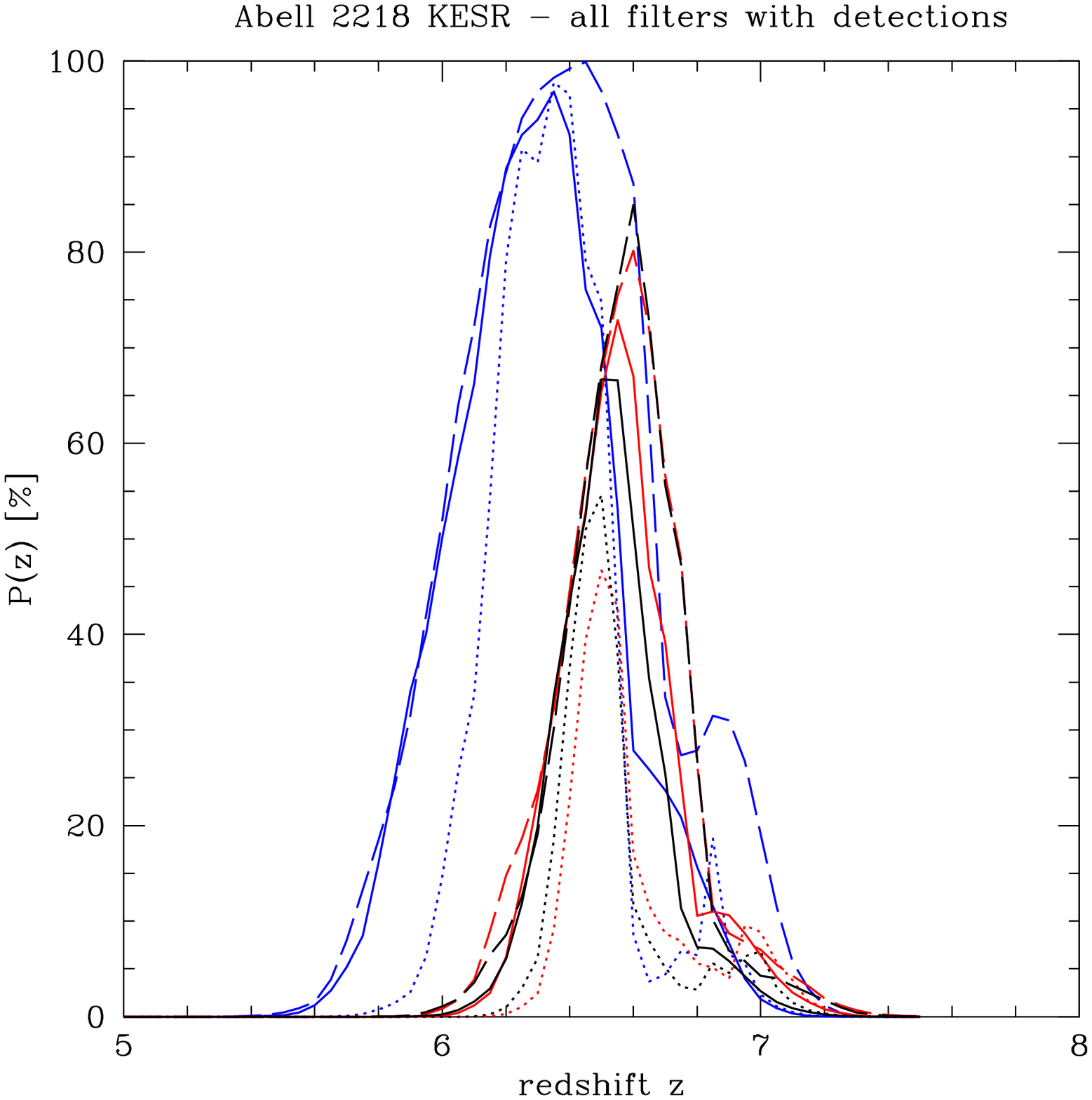,width=7cm}
	    \psfig{figure=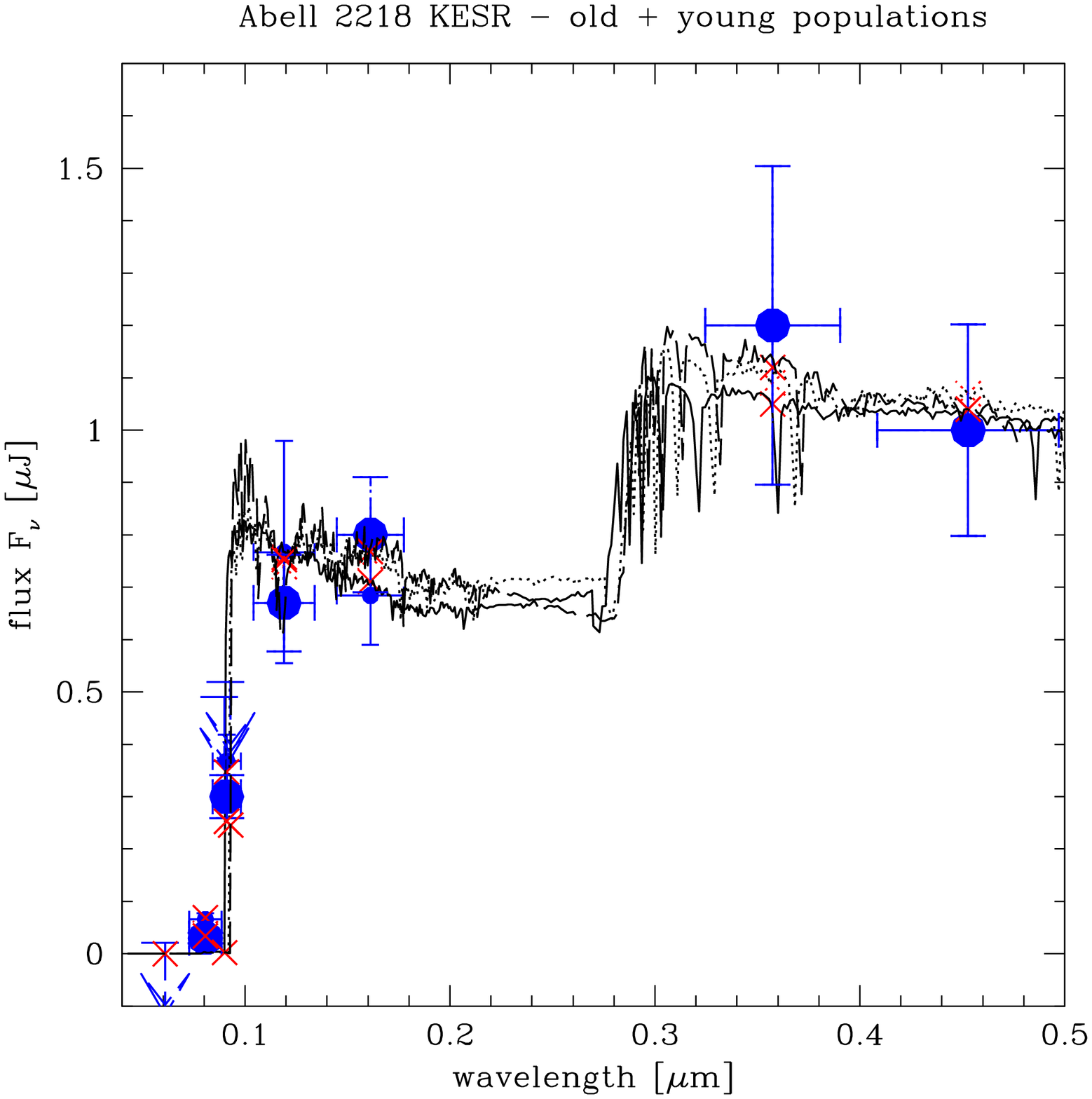,width=7cm}}
\caption{{\bf Left:} 
Photometric redshift probability
distributions \pz\ of Abell 2218 KESR using
three spectral template groups (solid, dotted, long dashed),
%Solid line: BCCWW template group, dotted: SB+QSO, long dashed: S03+. 
three different variants of the SED, and the photometry from all filters 
in which the object is detected (\iwfpc\ to 4.5 \micron)
assuming a minimum photometric error of 0.15 mag.
{\bf Right:} Best fits SEDs to the observations of Abell 2218
from Egami \etal\ 2004 (SED2)
%, including the flux limit from the
%non-detections in \vwfpc\ and at 9000-9300 \AA\ from spectroscopy;
%cf.\ Sect.\ \ref{s_obs}).
The red crosses indicate the corresponding model broad band fluxes.
The solid line shows the best fit for a template from the S03+ group,
and dotted from the SB+QSO group.
The redshift for these solutions are $z \sim$ 6.63 and 6.54 respectively.
See text for more information}
\label{fig_7}
\end{figure*}

As a spectroscopic redshift has not been obtained (yet) for this galaxy
we here examine its photometric redshift estimate.
In Fig.\ \ref{fig_7} (left) we show the photometric redshift probability
distributions \pz\ for the three SEDs (SED1-3) of Abell 2218 KESR
using three spectral template groups and adopting a minimum photometric error
of 0.15 mag.
For each redshift, \pz\ quantifies the quality of the best fit model
obtained varying all other parameters (i.e.\ extinction, \flyf, spectral template amoung
template group).
Given the excellent HST (WFPC2, ACS and NICMOS) photometry, \pz\
is quite well defined: the photometric redshift ranges typically between 
$z_{\rm phot} \sim$ 5.5 and 7.3. Outside of the plotted redshift range
\pz\ is essentially zero.

To summarise (but cf.\ Schaerer \& Pell\'o 2004), 
given the absence of a spectroscopic redshift, a fair number of 
good fits is found to the observations of Abell 2218 KESR
when considering all the free parameters. 
Three of them are illustrated in Fig.\ \ref{fig_7} (right). 
The main conclusions from these ``best fits'' are:
\begin{itemize}
\item[{\em 1)}] Generally the determined extinction is negligible or zero
	quite independently of the adopted extinction law.
	For few empirical templates we find good fits requiring an additional
	$A_V \sim$ 0.2--0.6 mag, depending on the adopted extinction law.
\item[{\em 2)}] Although generally burst models fit somewhat better
	than those with constant star formation among the theoretical
	templates,	% (BC, S03+), 
	the data does not strongly constrain the star formation history.
\item[{\em 3)}] Typical ages between $\sim$ 15 and 400 Myr are obtained.
	A reasonable 1-$\sigma$ upper bound on the age of $\sim$ 650 Myr can be 
	obtained assuming constant star formation.
	Young solutions ($\sim$ 15 and even younger) are obtained with burst models 
	or some empirical templates.
	The relatively modest strength  Balmer break observed between the HST and Spitzer 
	broad-band photometry does not necessarily imply old ages.
\item[{\em 4)}] Given degeneracies of the restframe UV spectra between age
  	and metallicity (cf.\ above) no clear indication on the galaxian
	metallicity can be derived, in contrast to the claim of KESR.
	Good fits to the available data can even be 
	found with solar metallicity starburst templates.
\item[{\em 5)}] Depending on the star formation history and age
	one may or may not expect intrinsic \lya\ emission, i.e.\
	an important \hii\ region around the object.
	The apparent absence of observed \lya\ emission does therefore
	not provide much insight.
\end{itemize} 

The theoretical templates can also be used to estimate the 
stellar mass involved in the starburst or the star formation
rate when constant star formation is assumed. For this aim 
%we use all the best fits to the three SEDs (SED1-3)
%with the S03+ templates, 
we assume a typical redshift of 
$z=6.6$, and the magnification $\mu=25$ determined by KESR.
%For the adopted cosmology the distance luminosity is then
%$d_L=$64457.8 Mpc.
% ==> 4*pi*d_L^2 = 4.79e+59 cm^2
%
For constant SF we obtain
%is assumed one obtains the following star formation rate: 
$SFR \sim (0.9-1.1)$ \msunyr\
(for a Salpeter IMF from 1 to 100 \msun). 
% b ~ (2.16-2.79)e-12
For the best fit ages of $\sim$ 400--570 Myr the total mass 
of stars formed would then correspond to $\sim (3.6-6.3) \times 10^8$ \msun.
The mass estimated from best fit burst models (of ages $\sim$ 6--20 Myr) is
slightly smaller, $M_\star \sim (0.3 - 1) \times 10^8$ \msun.
% b ~ 8.5e-5 to 2.5e-4
%
If we assume a Salpeter
IMF with \mlow\ $=0.1$ \msun\ the mass and SFR estimates would be higher 
by a factor 2.55, and in good agreement with the values derived by KESR 
and Egami \etal.
In all the above cases the total luminosity (unlensed) is typically $L_{\rm bol}
\sim 2 \times 10^{10}$ \lsun.

%
%%%%%%%%%%%%%%%%%%%%%%%%%%%%%%%%%%%%%%%%%%%%%%%%%%%%%%%%%%%%%%%%%%%%%
\section{$z \protect\ga 6$ starbursts: with Herschel and ALMA,
and now $\ldots$}

Let us now assume that starburst galaxies with dust exist at $z \ga 6$
and briefly examine their observability with facilities such as
Herschel and ALMA.
To do so we must assume a typical galaxy spectrum including the dust emission. 
For simplicity we here adopt the SED model by Melchior et al.\ (2001)
based on PEGASE.2 stellar modeling, on the D\'esert et al.\ (1990) dust
model, and including also synchrotron emission.
Their predicted SED for a galaxy with an SFR and/or total luminosity
quite similar to that estimated above for HCM6A is shown in Fig.\ 3.

\begin{figure}[htb]
%\centerline{\psfig{figure=melchiora_fig3.eps,width=7.5cm}}
\centerline{\psfig{figure=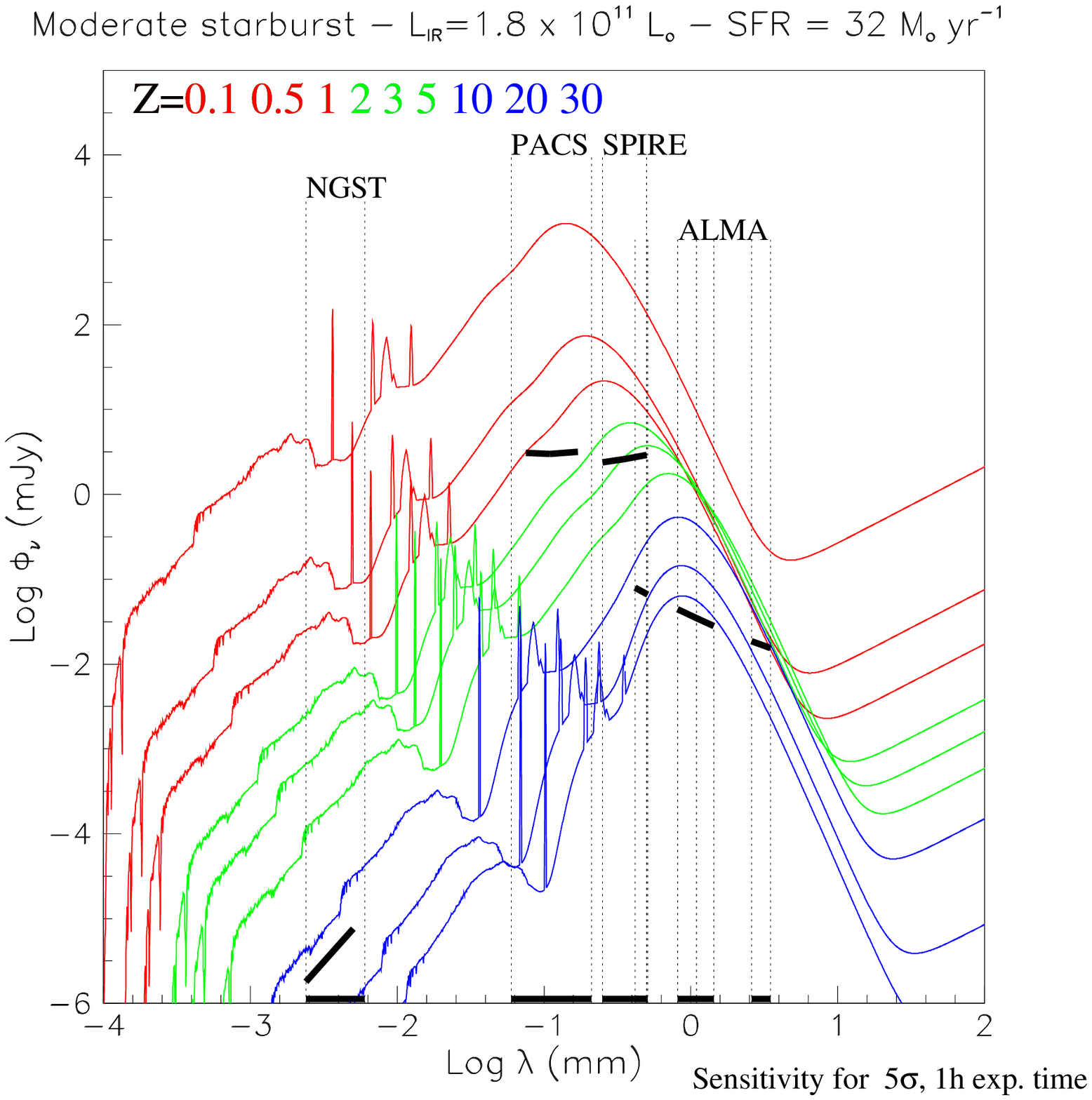,width=7.5cm}}
\caption{Predicted spectrum for a ``moderate'' starburst with SFR=32 \msunyr\
and $L \sim 1.8 \times 10^{11} \protect\lsun$} placed at redshift $z=$0.1, 0.5, 1, 2, 3, 5, 10, 20,
and 30. The thresholds of the JWST (here NGST), PACS and SPIRE onboard Herschel,
and ALMA are also presented. Figure taken from Melchior et al.\ (2001) with kind permission
\label{}
\end{figure}

Figure 3 shows the exquisite sensitivity of ALMA in the various bands
allowing in principle an easy detection of such objects up to redshift $\sim$ 10 
or even higher!

On the other hand, with the sensitivity of PACS and SPIRE on Herschel blank field 
observations of such an object are limited to smaller redshift ($z \la$ 1--4).
However, already with a source magnification of $\mu \sim$ 3--10 or more the
``template galaxy'' shown in Fig.\ 3 becomes observable with SPIRE
at $\sim$ 200--670 \micron.
In fact, such magnifications (and even higher ones) are not exceptional
in the central parts of massive lensing clusters. E.g.\ in our near-IR search 
conducted in two ISAAC fields ($\sim$ 2.5x2.5 arcmin$^2$) of two lensing clusters,
a fair number ($\sim$ 10--20) of $z \ga$ 6--7 galaxy candidates with $\mu \ga 5$
are found (Pell\'o et al.\ 2004b and these proceedings, Richard et al.\ 2004, in preparation). 
More than half of them have actually magnifications $\mu \ga 10$.
Such simple estimates show already quite clearly the potential of
strong gravitational lensing to extend the horizon of SPIRE/Herschel observations
beyond redshift $z \ga 5$!

Obviously a more rigorous feasibility study must also address the following
issues: 
How frequent is dust present in high-z galaxies? and up to what redshift?
We now have some indications for dust in one lensed $z=6.56$ galaxy (see Section 2)
and of course in high-z quasars. But how general/frequent is this?
How typical is the SED adopted above? The long wavelength emission due to dust
depends on various parameters such as metallicity, the dust/gas ratio, geometry,
the ISM pressure etc. 
Furthermore spatial resolution and source confusion are key issues which
must be addressed and which should vary quite strongly between blank fields
and cluster environments.
Last, but not least, the field of view of the various instruments is determinant
for the efficiency with which high-z candidates can be found and studied.
Several of these issues have already been partly addressed earlier
(cf.\ the 2000 Herschel conference proceedings of Pilbratt et al.\ 2001,
also Blain et al.\ 2002).

It is evident that various ground-based and space bourne facilities and
instruments will be used together to provide an optimal coverage
in wavelength, spatial resolution and field size, and to obtain imaging
as well as spectroscopy. 
%The use of gravitational lensing 
Near-IR wide field imagers and near-IR multi-object spectrographs
on 8-10m class telescopes and later with ELTs will undoubtably ``team up''
with the JWST, Herschel and ALMA to explore the first galaxies
in the Universe and their evolution from the Dark Ages to Cosmic Reionisation.
The wonderful power offered by gravitational lensing will continue to 
provide deeper or ``enhanced'' views of prime interest for the exploration
of the early Universe.

%%%%%%%%%%%%%%%%%%%%%%%%%%%%%%%%%

\end{document}